\newcommand{\diracslash}[1]{#1\llap{/\kern2pt}}

\newcommand{\be}{\begin{equation}}
\newcommand{\ee}{\end{equation}}
\newcommand{\bea}{\begin{eqnarray}}
\newcommand{\eea}{\end{eqnarray}}
\newcommand{\ba}[1]{\begin{array}{#1}}
\newcommand{\ea}{\end{array}}

\newcommand{\bt}{\begin{tabular}}
\newcommand{\et}{\end{tabular}}
\newcommand{\Tr}{{\rm Tr}}

\newcommand{\pa}{\partial}
\newcommand{\beas}{\begin{eqnarray*}}
\newcommand{\eeas}{\end{eqnarray*}}

\newcommand{\fr}{\frac}

\newcommand{\dg}{\dagger}

\newcommand{\pam}{\partial_\mu}

\documentclass[preprint,prd,aps,floats,nofootinbib,floatfix]{revtex4}
\usepackage{graphicx}

\begin{document}

\title{Kaons and antikaons in strong magnetic fields} 
\author{Amruta Mishra}
\email{amruta@physics.iitd.ac.in}
\affiliation{Department of Physics, Indian Institute of Technology, Delhi,
Hauz Khas, New Delhi -- 110 016, India}

\author{Anuj Kumar Singh}
\email{anuj0630@gmail.com}
\affiliation{Department of Physics, Indian Institute of Technology, Delhi,
Hauz Khas, New Delhi -- 110 016, India}

\author{Neeraj Singh Rawat}
\email{neerajrawat377@gmail.com}
\affiliation{Department of Physics, Indian Institute of Technology, Delhi,
Hauz Khas, New Delhi -- 110 016, India}

\author{Pratik Aman}
\email{pratikaman.atom@gmail.com}
\affiliation{Department of Physics, Indian Institute of Technology, Delhi,
Hauz Khas, New Delhi -- 110 016, India}

\begin{abstract}
The in-medium masses of the kaons and antikaons 
in strongly magnetized asymmetric nuclear matter are studied
using a chiral SU(3) model.
The medium modifications of the masses of these open strange
pseudoscalar mesons arise due to their interactions with the
nucleons and scalar mesons within the model. 
The effects of anomalous magnetic moments (AMM) of the nucleons
are taken into consideration in the present study and these
are seen to be large at high magnetic fields and
high densities. The isospin effects are appreciable
at high densities. The density effects are observed 
to be the dominant medium effects, as compared to
the effects from magnetic field and isospin asymmetry.
\end{abstract}
\maketitle

\def\bfm#1{\mbox{\boldmath $#1$}}

\section{Introduction}
The topic of hadrons in hot and/or dense matter has been 
a subject of extensive research due to its relevance 
in relativistic heavy ion collision experiments
in various high energy particle accelerators, as well
as, for study of bulk properties of astrophysical objects,
e.g., neutron stars. There are huge magnetic fields
produced in noncentral ultrarelativistic heavy ion
collision experiments 
\cite{HIC_mag_1,HIC_mag_2,HIC_mag_3,HIC_mag_4}. 
There is also natural occurrence
of large magnetic fields in magnetars. 
The study of the effects of magnetic fields, in addition to 
the effects of high density and/or temperature 
on the properties of hadrons has emerged as an intense
research topic in strong interaction physics  
in the recent years.
The initial system of the heavy ion collisions has
isospin asymmetry (as $N\ne Z$ in the colliding nuclei)
and hence it is important to study the effects of the
isospin asymmetry on the properties of the hadrons.
The various approaches used for the study of in-medium
properties of hadrons are the QCD sum rule approach
\cite{svznpb1,svznpb2}, the effective hadronic models,
e.g., Quantum Hadrodynamics (QHD) model \cite{walecka_1,walecka_2} 
and its generalizations, the Quark meson coupling (QMC)
model \cite{qmc_1,qmc_2} where the quarks/antiquarks inside of the hadron
interact via exchange of scalar and vector mesons,
and effective hadronic models based on low energy properties 
of QCD, e.g., chiral symmetry breaking, as well as,
coupled channel approach 
\cite{coupled_channel_1,coupled_channel_2,coupled_channel_3,coupled_channel_4,coupled_channel_5,coupled_channel_6}.

In the present work, using a chiral SU(3) model \cite{papa,kristof1},
the medium modifications of the masses of the kaons and antikaons 
in asymmetric nuclear matter in the presence of strong magnetic 
fields are studied. 
The proton has contributions from the Landau energy levels 
in the presence of an external magnetic field. 
The medium modifications of kaons and antikaons
arise due to the interaction with the nucleons 
and scalar mesons in the magnetized nuclear matter, 
within the chiral SU(3) model. 
The masses of the charged $K^\pm$ mesons have additional 
positive shift in the magnetized medium, arising from
the direct interaction of these mesons with the magnetic field.
The model has been used to study,
at zero magnetic field, the kaons and antikaons 
in isospin asymmetric nuclear (hyperonic) matter 
\cite{kristof1,kaon_antikaon,isoamss,isoamss1,isoamss2},
finite nuclei \cite{papa}, vector mesons \cite{vecm,am_vecqsr}, 
neutron stars \cite{nstar}.
The in-medium masses of open (strange) charm 
\cite{amdmeson,amarindamprc,amarvdmesonTprc,amarvepja,DP_AM_Ds},
open bottom mesons \cite{DP_AM_bbar,DP_AM_Bs},
charmonium \cite{amarvepja,jpsi_etac_AMarv} and 
bottomonium states \cite{AM_DP_upsilon}
have also been studied within the hadronic model
by generalizing the model to the charm and bottom
sectors. 
The partial decay widths of the charmonium states to $D\bar D$ pair
\cite{3p0_1,3p0_2,3p0_3,3p0_4,friman,amarvepja,amspmwg}, as well as,
of bottomonium states to $B\bar B$ \cite{amspm_upsilon}
in the hadronic medium have been studied, from the mass
modifications of these hidden and open heavy flavour mesons.
Recently, the effects of magnetic fields have been considered
on the light vector mesons \cite{vecqsr_mag} as well as 
the heavy flavour mesons 
\cite{dmeson_mag,bmeson_mag,charmonium_mag,upsilon_mag,jpsi_etac_mag}, 
through the coupling of the nucleons with the uniform magnetic
field, within the chiral effective model. The effects of the
anomalous magnetic moments of the nucleons
\cite{broderick1,broderick2,Wei,mao,amm,VD_SS_1,VD_SS_2,aguirre_fermion}
on the mass modifications of these hadrons are also studied.
 
%
The outline of the paper is as follows : In section II, we describe
briefly the chiral $SU(3)$ model used to investigate the mass
modifications of the kaons and antikaons in asymmetric nuclear
matter in the presence of strong magnetic fields. 
Section III discusses the results of the medium modifications
of the masses of these open strange pseudoscalar mesons 
in the magnetized  asymmetric nuclear matter.
In section IV, we summarize the findings of the present investigation.

\section{The chiral SU(3) model }
We use a chiral $SU(3)$ model \cite{papa} to obtain the in-medium
masses of the kaons and antikaons in magnetized nuclear matter.
The model is based on the nonlinear realization of chiral 
symmetry \cite{weinberg,coleman_1,coleman_2,bardeen} 
and broken scale invariance \cite{papa,kristof1,vecm}. 
The concept of broken scale invariance leading to the trace anomaly 
in QCD, $\theta_{\mu}^{\mu} = \frac{\beta_{QCD}}{2g} 
{G^a}_{\mu\nu} G^{\mu\nu a}$, where $G_{\mu\nu}^{a} $ is the 
gluon field strength tensor of QCD, is simulated in the effective 
Lagrangian at tree level through the introduction of 
a logarithmic potential in terms of a scalar dilaton field, $\chi$
\cite{sche1,ellis_1,ellis_2}.

The Lagrangian density of the chiral SU(3) model, in the presence
of magnetic field, is given as \cite{papa,dmeson_mag,bmeson_mag}
\bea
{\cal L}  = {\cal L}_{kin} + \sum_{ W =X,Y,V,{\cal A},u }{\cal L}_{BW}
          +  {\cal L}_{vec} + {\cal L}_0 +
{\cal L}_{scalebreak}+ {\cal L}_{SB}+{\cal L}_{mag}^{B\gamma},
\label{genlag} \eea
where, $ {\cal L}_{kin} $ corresponds to the kinetic energy terms
of the baryons and the mesons,
${\cal L}_{BW}$ contains the baryon-meson interactions,
$ {\cal L}_{vec} $ describes the dynamical mass
generation of the vector mesons via couplings to the scalar fields
and contains additionally quartic self-interactions of the vector
fields, ${\cal L}_0 $ contains the meson-meson interaction terms
${\cal L}_{scalebreak}$ is a scale invariance breaking logarithmic
potential, $ {\cal L}_{SB} $ describes the explicit chiral symmetry
breaking and ${\cal L}_{mag}^{B\gamma}$ describes the effects 
due to the magnetic field through the coupling of the baryons to 
the electromagnetic field.
The kinetic energy terms are given as
\bea
\label{kinetic}
{\cal L}_{kin} &=& i\Tr \overline{B} \gamma_{\mu} D^{\mu}B
                + \frac{1}{2} \Tr D_{\mu} X D^{\mu} X
+  \Tr (u_{\mu} X u^{\mu}X +X u_{\mu} u^{\mu} X)
                + \frac{1}{2}\Tr D_{\mu} Y D^{\mu} Y 
 \nonumber \\
               &+&\frac {1}{2} D_{\mu} \chi D^{\mu} \chi
                - \frac{ 1 }{ 4 } \Tr
\left(\tilde V_{ \mu \nu } \tilde V^{\mu \nu }  \right)
- \frac{ 1 }{ 4 } \Tr \left( {\cal A}_{ \mu \nu } {\cal A}^{\mu \nu }
 \right)
- \frac{ 1 }{ 4 } \Tr \left(F_{ \mu \nu } F^{\mu \nu }  \right),
\eea
where, $B$ is the baryon octet, $X$ is the scalar meson
multiplet, $Y$ is the pseudoscalar chiral singlet, 
$\chi$ is the scalar dilaton field,
$\tilde{V}_{\mu\nu}=\pa_\mu\tilde{V}_\nu-\pa_\nu\tilde{V}_\mu$,
${\cal A}_{\mu\nu}= \pa_\mu{\cal A}_\nu-\pa_\nu{\cal A}_\mu $, 
and $F_{\mu\nu}= \pa_\mu A_\nu-\pa_\nu A_\mu $, 
are the field strength tensors of 
the renormalised vector meson multiplet, $\tilde{V}^\mu$, 
the axial vector meson multiplet ${\cal A}^\mu$ 
and the photon field, $A^\mu$.
In Eq. (\ref{kinetic}), 
$u_\mu= -\fr{i}{4} [(u^\dg(\pam u)-(\pam u^\dg) u) 
 - (u (\pam u^\dg)-(\pam u) u^\dg)]$,
where, $u=\exp\Bigg[\fr{i}{\sigma_0}\pi^a\lambda^a\gamma_5\Bigg]$.
The covariant derivative of a field $\Phi (\equiv B,X,Y,\chi)$
reads $ D_\mu \Phi = \pam\Phi + [\Gamma_\mu,\Phi]$, with
$\Gamma_\mu=-\fr{i}{4} [(u^\dg(\pam u)-(\pam u^\dg) u)
 + (u (\pam u^\dg)-(\pam u) u^\dg)]$.
The interaction of the baryons with the meson, $W$ (scalar,
pesudoscalar, vector, axialvector meson) given by ${\cal L}_{BW}$,
as well as the terms ${\cal L}_{vec}$, ${\cal L}_0$,
${\cal L}_{scalebreak}$, ${\cal L}_{SB}$ have been 
described in detail in Ref. \cite{papa}. The term
${\cal L}_{mag}^{B\gamma}$, describing the interacion
of the baryons with the electromagnetic field
is given as 
\cite{dmeson_mag,bmeson_mag}
\be 
{\cal L}_{mag}^{B\gamma}=-{\bar {\psi_i}}q_i 
\gamma_\mu A^\mu \psi_i
-\frac {1}{4} \kappa_i \mu_N {\bar {\psi_i}} \sigma ^{\mu \nu}F_{\mu \nu}
\psi_i,
\label{lmag_Bgamma}
\ee
where, $\psi_i$ corresponds to the $i$-th baryon.
The tensorial interaction of baryons 
with the electromagnetic field given by the second term 
in the above equation is related to the
anomalous magnetic moments of the baryons
\cite{broderick1,broderick2,Wei,mao,amm,VD_SS_1,VD_SS_2,aguirre_fermion}. 
We choose the magnetic field to be uniform and along the
z-axis, and take the vector potential to be
$A^\mu =(0,0,Bx,0)$. 

The calculations in the present work of study of the kaons and
antikaons in the magnetized nuclear matter are carried out in
the mean field approximation, where the meson fields are replaced
by their expectation values. 
In addition to using the mean field approximation,
where the meson  fields are replaced by their expectation
values, we also use the approximations that
$\bar \psi_i \psi_j = \delta_{ij} \langle \bar \psi_i \psi_i
\rangle \equiv \delta_{ij} \rho_i^s $
and 
$\bar \psi_i \gamma^\mu \psi_j = \delta_{ij} \delta^{\mu 0} 
\langle \bar \psi_i \gamma^ 0 \psi_i
\rangle \equiv \delta_{ij} \delta^{\mu 0} \rho_i $, 
where, $\rho_i^s$ and 
$\rho_i$ are the scalar and number density of fermion of species, 
$i$ (neutron and proton in the present investigation).
We use the frozen glueball approximation,
i.e., we fix $\chi=\chi_0$, the vacuum value of the dilaton
field. This is due to the reason that
the medium modification of the dilaton field in nuclear medium
is observed to be negligible as compared to the medium changes of the
scalar fields, $\sigma$, $\zeta$ and $\delta$.
The coupled equations of motions of these scalar fields 
are derived from the Lagrangian density of the chiral SU(3) model,
and, are solved to obtain the values of these fields
in the asymmetric nuclear medium in the presence of 
magnetic field. The number and scalar densities 
of the proton have contributions from the Landau energy levels
and the neutrons have contributions to their number and scalar densities
due to the anomalous magnetic moment, in the presence
of a magnetic field \cite{dmeson_mag,bmeson_mag}. 
The expresssions for the number and scalar densities of the proton
in the presence of a uniform magnetic field (chosen to be
along z-direction) and accounting for the anomalous magnetic moments
for the nucleons are given as \cite{Wei,mao,dmeson_mag,bmeson_mag}
\begin{equation}
\rho_p=\frac{eB}{4\pi^2} \Bigg [ 
\sum_{\nu=0}^{\nu_{max}^{(S=1)}} k_{f,\nu,1}^{(p)} 
+\sum_{\nu=1}^{\nu_{(max)}^{(S=-1)}} k_{f,\nu,-1}^{(p)} 
\Bigg]
\label{rhop_mag}
\end{equation}
and 
\begin{eqnarray}
\rho^s_p &=& \frac{eB{m_p^*}}{2\pi^2} \Bigg [ 
\sum_{\nu=0}^{\nu_{max}^{(S=1)}}
\frac {\sqrt {{m_p^*}^2+2eB\nu}+\Delta_p}{\sqrt {{m_p^*}^2+2eB\nu}}
\ln |\frac{ k_{f,\nu,1}^{(p)} + E_f^{(p)}}{\sqrt {{m_p^*}^2
+2eB\nu}+\Delta_p}|\nonumber \\
&+&\sum_{\nu=1}^{\nu_{max}^{(S=-1)}}
\frac {\sqrt {{m_p^*}^2+2eB\nu}-\Delta_p}{\sqrt {{m_p^*}^2+2eB\nu}}
\ln|\frac{ k_{f,\nu,-1}^{(p)}+E_f^{(p)}}{\sqrt {{m_p^*}^2
+2eB\nu}-\Delta_p}|\Bigg]\nonumber\\
\label{rhps_mag}
\end{eqnarray}
where, $k_{f,\nu,\pm 1}^{(p)}$ are the Fermi momenta of protons
for the Landau level, $\nu$ for the spin index, $S=\pm 1$,
i.e. for spin up and spin down projections for the proton.
These Fermi momenta are related to the Fermi energy of the
proton as
\begin{equation}
k_{f,\nu,S}^{(p)}=\sqrt { {E_f^{(p)}}^2
-\Big (
{\sqrt {{m_p^*}^2+2eB\nu}+S\Delta_p}\Big )^2}.
\label{kfp_mag}
\end{equation}
The number density and the scalar density of neutrons are given as
\begin{eqnarray}
\rho_{n}&= & \frac{1}{4\pi^2} \sum _{S=\pm 1}
\Bigg \{ \frac{2}{3} {k_{f,S}^{(n)}}^3
+S\Delta_n \Bigg[ (m_n^*+S\Delta_n) k_{f,S}^{(n)}
\nonumber \\
&+&
{E_f^{(n)}}^2 \Bigg( arcsin \Big (
\frac{m_n^*+S\Delta_n}{E_f^{(n)}}\Big)-\frac{\pi}{2}\Bigg)\Bigg]
\Bigg \}
\label{rhon_mag}
\end{eqnarray}
and
\begin{equation}
\rho^s_n =\frac{m_n^*}{4\pi^2} \sum _{S=\pm 1} 
\Bigg [ k_{f,S}^{(n)} E_f^{(n)} - 
(m_n^*+S\Delta_n)^2 \ln | \frac {k_{f,S}^{(n)}+ 
E_f^{(n)}}{m_n^*+S\Delta_n} | \Bigg].
\label{rhns_mag}
\end{equation}
The Fermi momentum, $k_{f,S}^{(n)}$ 
for the neutron with spin projection, S 
($S=\pm 1$ for the up (down) spin projection), 
is related to the Fermi energy for the 
neutron, $E_f^{(n)}$ as
\begin{equation}
k_{f,S}^{(n)}= \sqrt { {E_f^{(n)}}^2 -
(m_n^*+S\Delta_n)^2}.
\label{kfn_mag}
\end{equation}
In the equations (\ref{rhop_mag})-(\ref{kfn_mag},
the parameter $\Delta _{i}$ is related to the
to the anomalous magnetic moment for the nucleon, $i$
($i=p,n$) as
$\Delta_i =-\frac{1}{2} \kappa_i \mu_N B$,
where, $\kappa_i$, occurring in the second term
in the Lagrangian density given by Eq. (\ref{lmag_Bgamma}),
is the value of the gyromagnetic ratio of the nucleon
corresponding to the anomalous magnetic moment 
of the nucleon.
Using the scalar densities of the nucleons in the presence
of magnetic field, the values of the scalar fields, 
$\sigma$, $\zeta$ and $\delta$
are obtained by solving their coupled equations of motion,

The interaction Lagrangian modifying the masses of the $K$ 
and $\bar{K}$ mesons can be written as \cite{isoamss2}
\begin{eqnarray}
\mathcal{L}_{K}^{int}&=&-\frac{i}{4f_k^2}[(2\bar{p}\gamma^\mu p
+\bar{n}\gamma^\mu n)
(K^-(\partial_ \mu K^+)
-(\partial _ \mu K^-)K^+)
\nonumber \\ &+& (\bar{p}\gamma^\mu p+2\bar{n}\gamma^\mu n)
(\bar{K^0}(\partial_\mu K^0)
-(\partial_\mu\bar{K^0})K^0)]
\nonumber \\ &+& \frac{m_K^2}{2f_k^2}
\Big [ (\sigma+\sqrt{2}\zeta+\delta)(K^+K^-)
+  (\sigma+\sqrt{2}\zeta-\delta)(K^0\bar{K^0})\Big]\nonumber\\&-
&\frac{1}{f_k}[(\sigma+\sqrt{2}\zeta+\delta)(\partial_\mu K^+)
(\partial^ \mu {K^-})+(\sigma+\sqrt{2}\zeta-\delta)
(\partial^ \mu K^0)(\partial^ \mu \bar{K^0})]\nonumber\\&+
&\frac{d_1}{2f_k^2}[(\bar{p}p+\bar{n}n)((\partial_\mu K^+)
(\partial^ \mu K^-)+(\partial_\mu K^0)
(\partial^ \mu \bar{K^0}))]
\nonumber\\&+&\frac{d_2}{2f_k^2}[\bar{p}p
(\partial_\mu K^+)(\partial^ \mu K^-)
+\bar{n}n(\partial_\mu K^0)(\partial^ \mu \bar{K^0})].
\label{lk_int}
\end{eqnarray}
In the above, the first term is the vectorial Weinberg Tomozawa 
interaction term, which is the leading contribution in the
chiral perturbation theory. This term is attractive for $\bar K$
mesons but repulsive for $K$ mesons. The next to leading 
contributions are given by the scalar meson exchange term
(the second term) and the range term (the last three terms).
The parameters $d_1$ and $d_2$ in the last two range terms 
are determined to be $ 2.56/m_K $ and $ 0.73/m_K $ 
respectively \cite{isoamss2}, 
by fitting to the empirical values 
of the kaon-nucleon (KN) scattering lengths
for I=0 and I=1 channels \cite{barnes}. 

The Fourier transformations of the equations of motion 
for kaons (antikaons) obtained from the total Lagrangian
density 
$${\cal L}_K =(\partial _\mu {\bar K}) (\partial^\mu K)-
m_{K(\bar K)}^2 \bar K K +{\cal L}_K^{int}$$
lead to the dispersion relations,
\begin{equation}
-\omega^2 + |{\vec k}|^2 +m_{K(\bar K)}^2
-\Pi_{K(\bar{K})}(\omega, | \vec{k} | )=0
\label{dispkkbar}
\end{equation}
where $\Pi_{K(\bar k)}$ denotes the kaon (antikaon) self energy 
in the medium.
Explicitly, these self energies are given as 
\begin{eqnarray}
&&\Pi_{K}(\omega,|\vec{k}|)=-\frac{1}{4f_k^2}
[3(\rho_p+\rho_n)\pm(\rho_p-\rho_n)] \omega 
+\frac{m_k^2}{2f_k}(\sigma^\prime+\sqrt{2}\zeta^\prime\pm\delta^\prime)
\nonumber \\ &+&
\Bigg[-\frac{1}{f_k}(\sigma^\prime+\sqrt{2}\zeta^\prime\pm\delta^\prime)
+\frac{d_1}{2f_k^2}(\rho_p^s+\rho_n^s) 
+\frac{d_2}{4f_k^2}[(\rho_p^s+\rho_n^s)
\pm(\rho_p^s-\rho_n^s)]\Bigg](\omega^2- |{\vec k}|^2)\nonumber \\
\label{selfk}
\end{eqnarray}
where the $\pm$ signs refer to the $K^+$ and $K^0$ respectively,
and, 
\begin{eqnarray}
&&\Pi_{\bar{K}}(\omega,|\vec{k}|)=\frac{1}{4f_k^2}
[3(\rho_p+\rho_n)\pm(\rho_p-\rho_n)] \omega
+\frac{m_k^2}{2f_k}[\sigma^\prime+\sqrt{2}
\zeta^\prime\pm\delta^\prime] \nonumber \\ &+& 
\Bigg[-\frac{1}{f_k}
(\sigma^\prime+\sqrt{2}\zeta^\prime\pm\delta^\prime)
+\frac{d_1}{2f_k^2}(\rho_p^s+\rho_n^s)
+
\frac{d_2}{4f_k^2}[(\rho_p^s+\rho_n^s)
\pm(\rho_p^s-\rho_n^s)]\Bigg](\omega^2-|{\vec k}|^2)\nonumber \\
\label{selfkbar}
\end{eqnarray}
where the $ \pm $ signs refer to the $ K^- $ and 
$ \bar{K^0} $ respectively.
In the above, $\sigma '$, $\zeta '$ and $\delta '$ 
are the deviations of the expectation values of the scalar fields
from their vacuum expectation values. 
The masses for $K(\bar K$), 
$m_{K(\bar K)}^*=\omega(|\vec k|$=0)),
as modified due to their interactions with the 
nucleons and scalar mesons in the magnetized nuclear matter, 
are calculated from the
dispersion relations given by equation (\ref{dispkkbar}),
using the self energies for the kaons and antikaons given
by equations (\ref{selfk}) and (\ref{selfkbar}). 
These mass modifications of the kaons and antikaons 
do not take into account the direct interaction 
of the charged kaons and antikaons
with the electromagnetic field.

%
We consider the interaction of the medium modified
charged $K^\pm$ mesons 
(due to interactions with the nucleons and scalar mesons 
in the magnetized nuclear matter) 
with mass $m_{K^\pm}^*$, with the uniform
magnetic field, $B$ through a minimal interacion,
which leads to a further shift in the mass
of the charged kaon (antikaon).
In the presence of the uniform magnetic field, $B$, 
the mass squared of the spin zero charged kaon (antikaon), 
has summation of positive energy shifts of $(2n+1)eB$ 
from the Landau levels (labelled by $n \ge 0$)
\cite{Chernodub_1,Chernodub_2,Aguirre_pion}. 
Retaining only the lowest Landau level, i.e, $n=0$ level
\cite{dmeson_mag,bmeson_mag}, the effective mass of the
charged $K^\pm$ meson in the magnetized nuclear medium,
is given as
\begin{eqnarray}
m_{K^{\pm}}^{eff}=\sqrt{(m_{K^{\pm}}^*)^2+eB},
\label{meffkpkm}
\end{eqnarray}
where, $m_{K^\pm}^*$ is the mass of the $K^\pm$ meson
arising due to interaction with the medium modified nucleons and
scalar mesons, as has already been mentioned, is the solution 
for $\omega$ at $|\vec{k}|=0$ of the dispersion relation 
given by Eq. (\ref{dispkkbar}) for $K^\pm$ meson.

For the neutral kaons (antikaons), 
$K^0(\bar{K^0})$, the effective masses are given as
\begin{eqnarray}
m_{K^0(\bar{K^0})}^{eff}=m_{K^0({\bar{K^0})}}^*
\label{meffk0k0bar}
\end{eqnarray}
In equations (\ref{meffkpkm}) and (\ref{meffk0k0bar}),
$m_{K({\bar{K})}}^*$ are the solutions 
for $\omega$ at $|\vec{k}|=0$ of the dispersion relations 
given by Eq. (\ref{dispkkbar}).
\begin{figure}
\includegraphics[width=15cm,height=15cm]{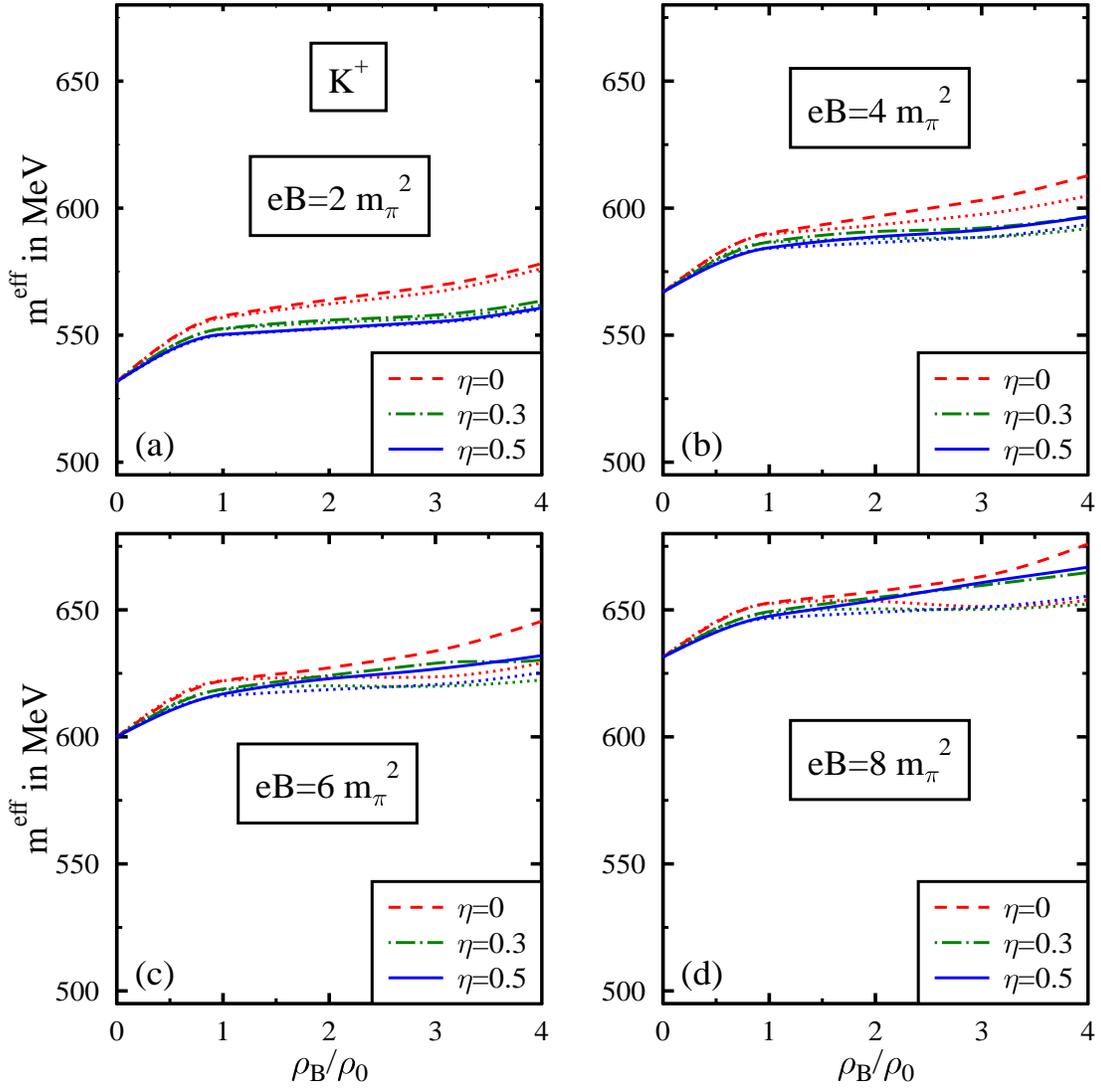}
\caption{(Color online)
Effective mass  of ${K^+}$ (in MeV) plotted as a function 
of the baryon density (in units of nuclear matter saturation density,
$\rho_0$), with different values of isospin asymmetric parameter 
$\eta$, accounting 
for the effects of the anomalous magnetic moments (AMMs) 
of the nucleons. The results are compared with the case 
of not accounting for anomalous magnetic moments 
(shown as dotted lines).}
\label{mkp_mag}
\end{figure}
\begin{figure}
\includegraphics[width=15cm,height=15cm]{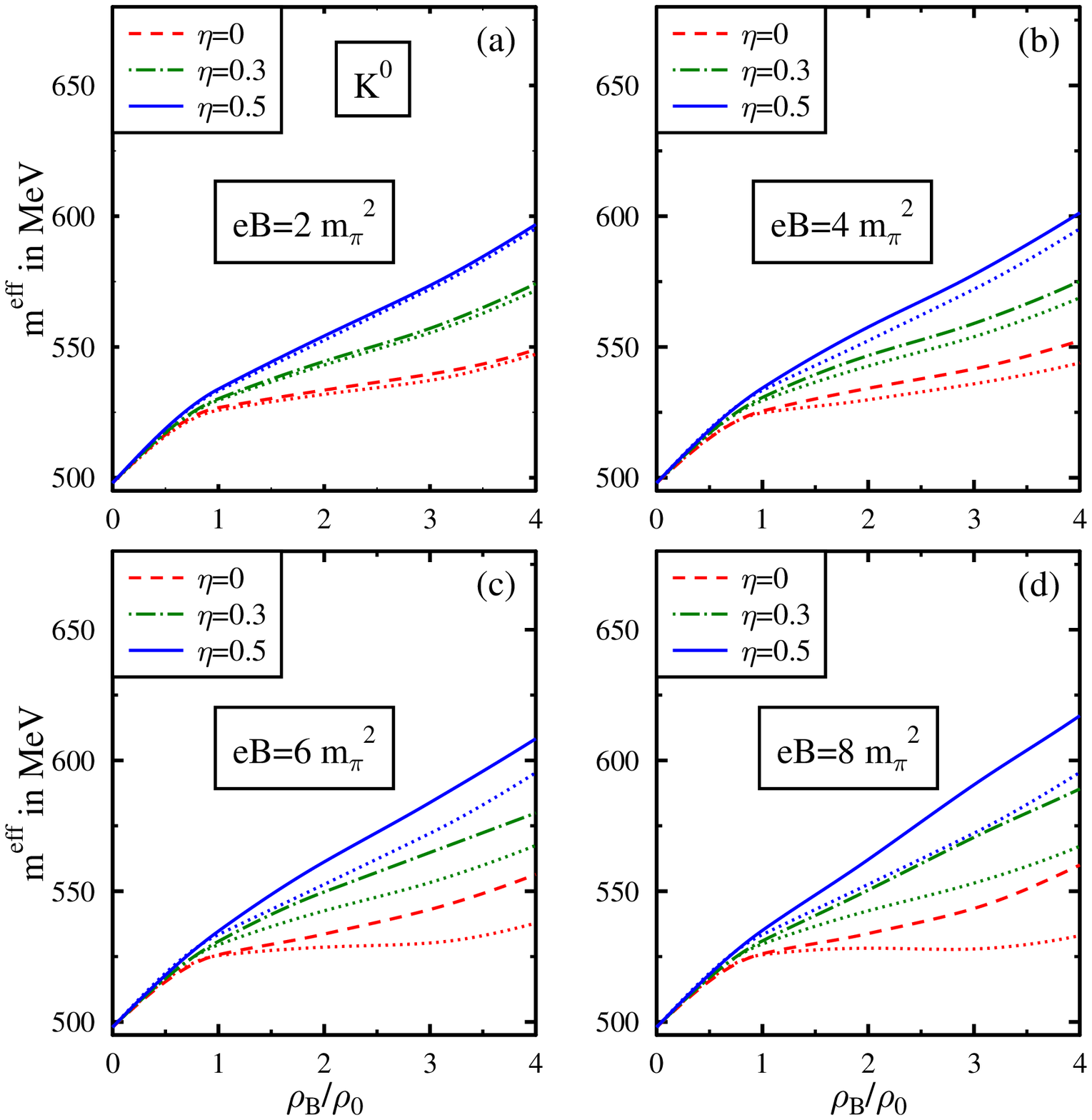}
\caption{(Color online)
Effective mass of $K^0$ (in MeV) plotted as a function 
of the baryon density (in units of nuclear matter saturation density,
$\rho_0$), with different 
values of isospin asymmetric parameter $\eta$ , accounting 
for the effects of the anomalous magnetic moments (AMMs) 
of the nucleons. The results are compared with the case 
of not accounting for anomalous magnetic moments 
(shown as dotted lines).}
\label{mk0_mag}
\end{figure}
\begin{figure}
\includegraphics[width=15cm,height=15cm]{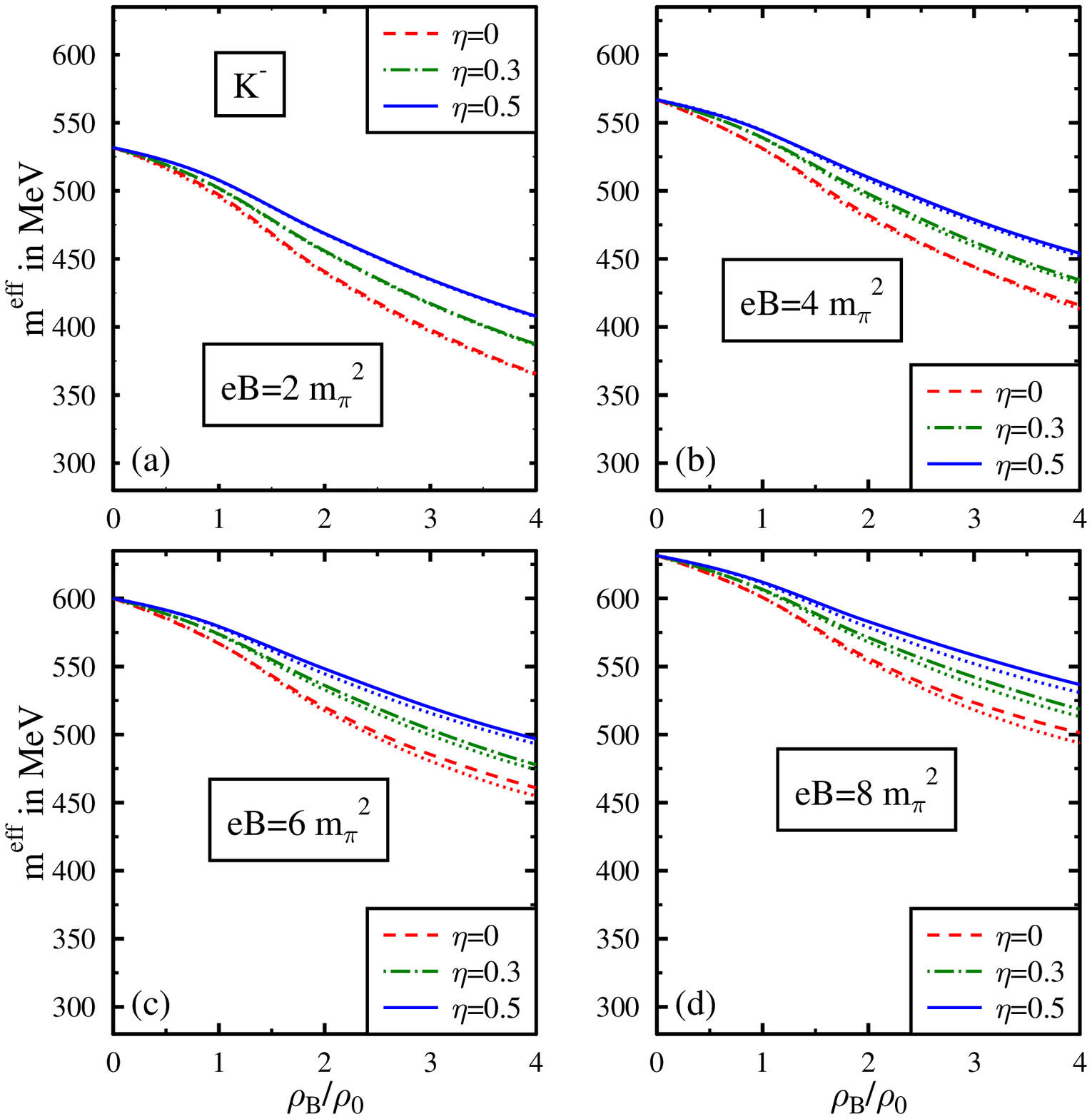}
\caption{(Color online)
Effective mass  of ${K^-}$ (in MeV) plotted as a function 
of the baryon density (in units of nuclear matter saturation density,
$\rho_0$), with different 
values of isospin asymmetric parameter $\eta$ , accounting 
for the effects of the anomalous magnetic moments (AMMs) 
of the nucleons. The results are compared with the case 
of not accounting for anomalous magnetic moments 
(shown as dotted lines).}
\label{mkm_mag}
\end{figure}
\begin{figure}
\includegraphics[width=15cm,height=15cm]{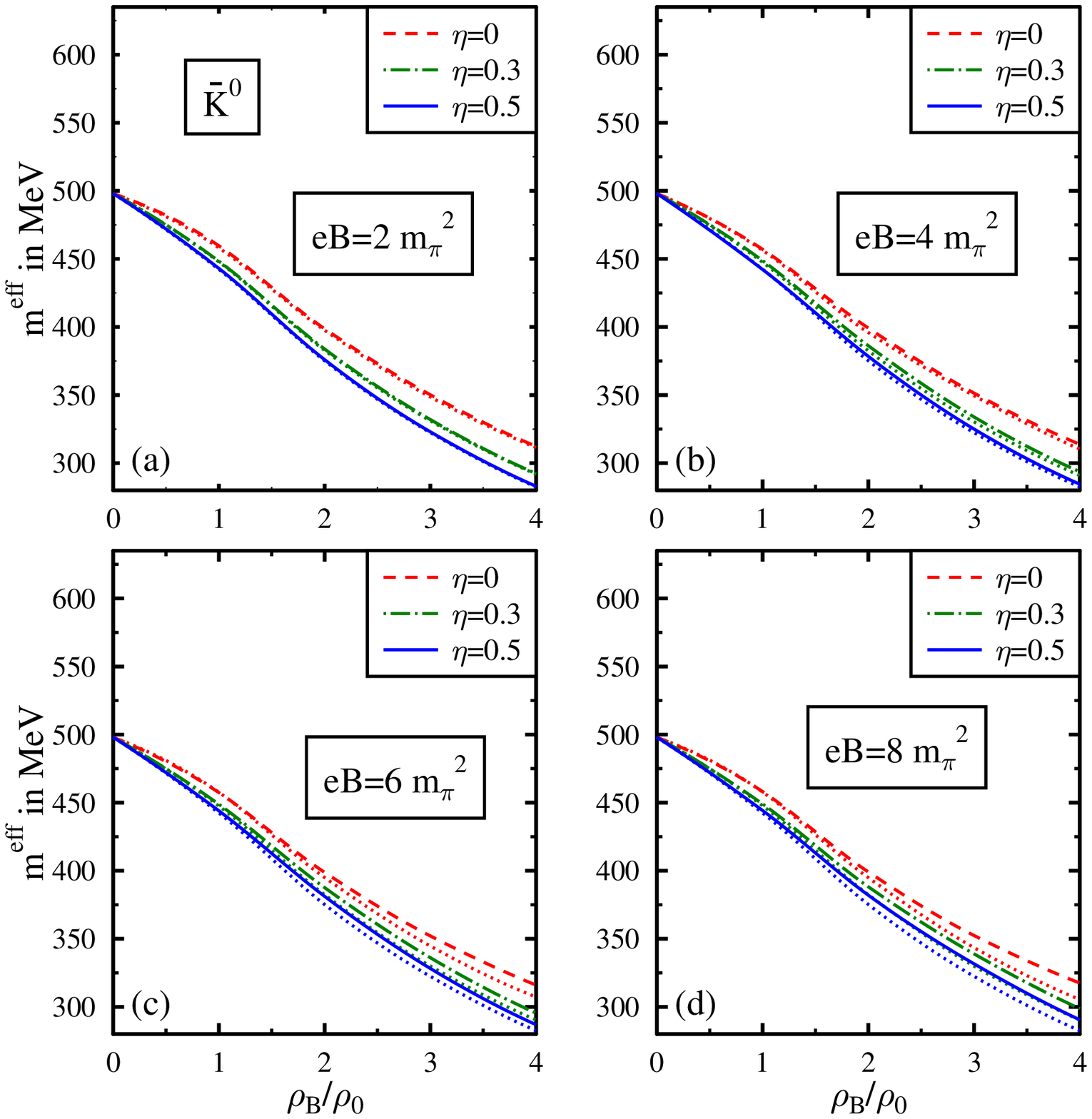}
\caption{(Color online)
Effective mass  of $\bar {K^0}$ (in MeV) plotted as a function 
of the baryon density (in units of nuclear matter saturation density,
$\rho_0$), with different 
values of isospin asymmetric parameter $\eta$ , accounting 
for the effects of the anomalous magnetic moments (AMMs) 
of the nucleons. The results are compared with the case 
of not accounting for anomalous magnetic moments 
(shown as dotted lines).}
\label{mk0bar_mag}
\end{figure}

\begin{table}
\begin{tabular}{|c|c|c|c|c|c|}
\hline
\multicolumn{2}{|c|}{
} & \multicolumn{2}{|c|}{$ \eta=0 $} 
&  \multicolumn{2}{|c|}{$ \eta=0.5 $}\\
\cline{3-6}
\multicolumn{2}{|c|}
 {${eB}/{m_\pi^2}$}
 &  $  \rho_B=\rho_0 $ & $ \rho_B=4\rho_0 $ & 
$ \rho_B=\rho_0 $ & $ \rho_B=4\rho_0 $ \\
\hline
\multicolumn{2}{|c|}{
{0}
}   & 
{521}  & 544.6    & 
{518}  & 
{528.9}  \\
\hline
{
2}& (a) 
&  556.9 & 576.2 & 
550.0  & 560.2  \\
\cline{2-6}
& (b) 
 & 557.6 & 578.2 & 
550.3 & 560.7        \\
\hline
{
4
}&  (a)
& 589.7 & 604.9 & 
584.2 &593.6 \\
\cline{2-6}
& (b)
& 590.2 & 612.8 & 
584.4 & 596.6 \\
\hline
{
6
}&  (a)
& 622.1 & 629.1 & 
616.2 &625.3\\
\cline{2-6}
& (b)
&622.1 & 645.5 & 
616.9 & 632.0\\
\hline
{
8}& (a)
& 652.5 & 653.8 & 
646.7 & 655.4\\
\cline{2-6}
& (b)
& 652.7 & 675.9 & 
647.5 & 666.8\\
\hline
\end{tabular}
\vskip 0.1in
\caption{In-medium masses for $K^+$ meson in magnetized nuclear
matter for densities of $\rho_0$ and 4$\rho_0$, asymmetric parameter,
$\eta$=0 and 0.5 and for magnetic fields, $eB/{m_\pi^2}$ as 2,4,6 and
8. These masses are compared with the in-medium masses
of $K^+$ meson for zero magnetic field.}
\label{table1}
\end{table}

\begin{table}
\begin{tabular}{|c|c|c|c|c|c|}
\hline
\multicolumn{2}{|c|}{
} 
& \multicolumn{2}{|c|}{$ \eta=0 $} 
&  \multicolumn{2}{|c|}{$ \eta=0.5 $}\\
\cline{3-6}
\multicolumn{2}{|c|}
 {${eB}/{m_\pi^2}$} 
&  $  \rho_B=\rho_0 $ & $ \rho_B=4\rho_0 $ & 
$ \rho_B=\rho_0 $ & $ \rho_B=4\rho_0 $ \\
\hline
\multicolumn{2}{|c|}{
{0}
}   & 
{524.8}  & 547.1   & 
{529}  & 
{591.5}  \\
\hline
{
2}& (a) 
& 525.8 & 547.2 
& 533.2 & 595.3  \\
\cline{2-6}
& (b) 
& 526.8 & 549.0 
& 533.9 & 596.8        \\
\hline
{
4
}&  (a)
& 524.8 & 543.9 
& 533.4 & 595.2  \\
\cline{2-6}
& (b)
& 525.5 & 552.4 
& 534.3 & 601.3  \\
\hline
{
6
}&  (a)
& 525.4 & 537.8 
& 533.4 & 595.3 \\
\cline{2-6}
& (b)
& 525.7 & 556.4 
& 534.7 & 608.3 \\
\hline
{
8}& (a)
& 525.7 & 533.0 
& 533.4 & 595.4 \\
\cline{2-6}
& (b)
& 526.1 & 560.0 
& 535.0 & 617.1 \\
\hline
\end{tabular}
\vskip 0.1in
\caption{In-medium masses for $K^0$ meson in magnetized nuclear
matter for densities of $\rho_0$ and 4$\rho_0$, asymmetric parameter,
$\eta$=0 and 0.5 and for magnetic fields, $eB/{m_\pi^2}$ as 2,4,6 and
8. These masses are compared with the in-medium masses
of $K^0$ meson for zero magnetic field.}
\label{table2}
\end{table}

\begin{table}
\begin{tabular}{|c|c|c|c|c|c|}
\hline
\multicolumn{2}{|c|}{
} &
  \multicolumn{2}{|c|}{$ \eta=0 $} 
&  \multicolumn{2}{|c|}{$ \eta=0.5 $}\\
\cline{3-6}
\multicolumn{2}{|c|} {${eB}/{m_\pi^2}$}
&  $  \rho_B=\rho_0 $ & $ \rho_B=4\rho_0 $ & 
$ \rho_B=\rho_0 $ & $ \rho_B=4\rho_0 $ \\
\hline
\multicolumn{2}{|c|}{
{0}
}   & 
{452} & 307.9   & 
471.9  & 
360.2  \\
\hline
{
2}& (a) 
& 495.4 & 365 
& 507.5 & 407.4    \\
\cline{2-6}
& (b) 
& 497.0 & 365.8 
 & 507.8  & 407.9        \\
\hline
{
4
}&  (a)
& 530.7  & 413.4 
& 544.3  & 452.3   \\
\cline{2-6}
& (b)
& 531.1  & 416.0 
 & 544.1  & 454.1   \\
\hline
{
6
}&  (a)
& 567.0  & 455.0 
& 578.6  & 439.2  \\
\cline{2-6}
& (b)
& 566.8  & 460.9  
& 579.4  & 496.8 \\
\hline
{
8}& (a)
& 600.7  & 493.9 
 & 611.0  & 530.9  \\
\cline{2-6}
& (b)
& 600.7  & 501.3 
 & 611.9  & 536.7  \\
\hline
\end{tabular}
\vskip 0.1in
\caption{In-medium masses for $K^-$ meson in magnetized nuclear
matter for densities of $\rho_0$ and 4$\rho_0$, asymmetric parameter,
$\eta$=0 and 0.5 and for magnetic fields, $eB/{m_\pi^2}$ as 2,4,6 and
8. These masses are compared with the in-medium masses
of $K^-$ meson for zero magnetic field.}
\label{table3}
\end{table}

\begin{table}
\begin{tabular}{|c|c|c|c|c|c|}
\hline
\multicolumn{2}{|c|}{
} & \multicolumn{2}{|c|}{$ \eta=0 $} 
&  \multicolumn{2}{|c|}{$ \eta=0.5 $}\\
\cline{3-6}
\multicolumn{2}{|c|}
 {${eB}/{m_\pi^2}$}
 &  $  \rho_B=\rho_0 $ & $ \rho_B=4\rho_0 $ & 
$ \rho_B=\rho_0 $ & $ \rho_B=4\rho_0 $ \\
\hline
\multicolumn{2}{|c|}{
{0}
}   & 
{455.9} & 310.9    & 
437.9  & 
278.6  \\
\hline
{
2}& (a) 
& 458.1  & 311.4 
 & 442.3  & 282.5    \\
\cline{2-6}
& (b) 
& 459.6  & 312.4 
 & 443.0  & 283.0         \\
\hline
{
4
}&  (a)
& 456.3  & 310.3 
 & 442.4  & 282.5  \\
\cline{2-6}
& (b)
& 457.0  & 313.8 
 & 442.2  & 284.5  \\
\hline
{
6
}&  (a)
& 457.6  & 307.3 
 & 442.4  & 282.5  \\
\cline{2-6}
& (b)
& 457.4  & 316.0 
& 444.0  & 286.9 \\
\hline
{
8}& (a)
& 457.8  & 305.4 
 & 442.4  & 282.5  \\
\cline{2-6}
& (b)
& 458.1  & 317.5 
& 444.2  & 290.5  \\
\hline
\end{tabular}
\vskip 0.1in
\caption{In-medium masses for $\bar {K^0}$ meson in magnetized nuclear
matter for densities of $\rho_0$ and 4$\rho_0$, asymmetric parameter,
$\eta$=0 and 0.5 and for magnetic fields, $eB/{m_\pi^2}$ as 2,4,6 and
8. These masses are compared with the in-medium masses
of $\bar {K^0}$ meson for zero magnetic field.}
\label{table4}
\end{table}

\section{Results and Discussions}
The in-medium masses of the $K$ and $\bar{K}$ mesons are investigated 
in asymmetric nuclear matter in the presence of an external magnetic 
field, using a chiral SU(3) model. The medium modifications 
arise due to the interactions of these mesons
with the protons, neutrons, and the scalar mesons  ($\sigma$, $\zeta$,
and $\delta$). The charged $K$ and $\bar{K}$ mesons, 
i.e, $K^{\pm}$ mesons
have additional mass modifications due to the Landau 
quantization effects in the presence of the external 
magnetic field, as given by Eq.(\ref{meffkpkm}).
In the presence of an external magnetic field 
$\vec{B}=(0,0,B)$, the values of the scalar fields, 
$\sigma$, $\zeta$ and $\delta$ in the magnetized nuclear matter 
are solved from their equations of motion, for given values
of the baryon density, $\rho_B$, the isospin
asymmetry parameter, $\eta=(\rho_n-\rho_p)/(2\rho_B)$
and the magnetic field, $B$.
These coupled equations of motion are obtained
in the mean field approximation (replacing the meson fields
by their mean values), along with the approximations  
$\bar \psi_i \psi_j = \delta_{ij} \langle \bar \psi_i \psi_i
\rangle \equiv \delta_{ij} \rho_i^s $ and 
$\bar \psi_i \gamma^\mu \psi_j = \delta_{ij} \delta^{\mu 0} 
\langle \bar \psi_i \gamma^ 0 \psi_i
\rangle \equiv \delta_{ij} \delta^{\mu 0} \rho_i $, 
where, $\rho_i^s$ and $\rho_i$ are the scalar and number density 
of the nucleons ($i$=p,n).
The number density and scalar density of the 
charged baryon, the proton, have contributions from the Landau 
energy levels in the presence of the external magnetic field. 
The in-medium masses of the $K$ and $\bar{K}$ 
mesons are calculated by using the dispersion relations 
given by (\ref{dispkkbar}), with the self-energies 
for the kaons $(K^+,K^0)$ and antikaons ($K^-,\bar{K^0})$ 
given by Eqs. (\ref{selfk}) and (\ref{selfkbar}) respectively. 
As has already been mentioned the charged $K^\pm$ mesons
have additional mass shifts due to direct interaction with the
electromagnetic field.

The effective massses of the kaons ($K^+$, $K^0$) 
and the antikaons ($K^-$ and $\bar{K^0}$) 
in the magnetized asymmetric nuclear matter,
are plotted as functions of $\rho_B/\rho_0$ 
(the baryon density in units of the nuclear matter saturation density), 
in Figs. \ref{mkp_mag}, \ref{mk0_mag}, 
\ref{mkm_mag} and \ref{mk0bar_mag}.
These are shown for various values of the isospin asymmetry parameter 
$ \eta $, including the effects from the anomalous 
magnetic moments (AMMs) of the nucleons. These are compared 
with the case that does not account for the effects 
from the anomalous magnetic moments (shown as dotted lines). 
At $\rho_B=0$, 
there are still mass modifications of charged kaons (antikaons), 
i.e, $K^{\pm}$ mesons, due to direct interaction with 
the electromagnetic field, as given by Eq. (\ref{meffkpkm}), 
while for the electrically charged neutral 
$K^0$ and $(\bar{K^0})$ mesons, there is no such 
mass modification due to magnetic field at $\rho_B=0$.
For the isospin asymmetric nuclear matter with
the isospin asymmetry parameter, $\eta=0.5$, 
the medium comprises of only neutrons, 
and hence the only effect of magnetic field is due to 
the anomalous magnetic moment (AMM) of the neutrons. 
Hence, in the case when the AMM effects are not 
taken into consideration, the expectaion values of the scalar fields
remain independent of the magnetic field.

For isospin symmetric nuclear matter ($\rho_p=\rho_n$), 
in the absence of a magnetic field, the scalar densities of
proton and neutron are identical and the value of the scalar isovector
meson, $\delta$ remains zero. The masses of the kaons and
antikaons are calculated from the dispersion relation
given by equation (\ref{dispkkbar}), with additional positive
shifts for the charged $K^\pm$, arising from the direct interaction
of these mesons with the magnetic field through minimal coupling.
The first term of the self energy,
the Weinberg Tomozawa term, leads to a rise (drop) in the mass
of the kaon (antikaon). The second term, which is the scalar 
exchange term is attractive for both the kaons and antikaons.
The last three terms in the self energy comprise the range
terms, of which, the first term is repulsive,
whereas the second and third range terms (the $d_1$ and $d_2$ terms)
are attractive for both kaons and antikaons.
It might be noted here that in the presence of a magnetic field,
due to the different interactions of the proton (charged nucleon)
and the neutron with the magnetic field, the value
of $\delta$ ($\sim (\rho_s^p-\rho_n^s$)) no longer remains 
zero. The coupled equations of motion for $\sigma$, $\zeta$
and $\delta$ are solved to obtain the values of these
fields (and hence the nucleon scalar densities), which are
used to calculate the masses of the kaons and antikaons
in the magnetized nuclear matter.
The contributions due to isospin asymmetry in the medium,
arise explicitly from the last terms of the Weinberg Tomozawa term,
the scalar exchange term, and the first and last range terms.
There is contribution from isospin asymmetry also 
from the second range term implicitly, as the values of
the scalar densities of the proton and neutron are different
in the isospin asymmetric medium (as calculated from the coupled
equations of motion for $\sigma$, $\zeta$ as well as $\delta$), 
as compared to in symmetric nuclear matter ($\eta=0$).

In isospin symmetric nuclear matter ($\eta$=0), as can be seen from
figures \ref{mkp_mag} and \ref{mk0_mag}, the masses of both $K^+$
as well as $K^0$ are observed to increase as  
baryon density is increased, with an initial rapid rise
upto a density of about nuclear matter
saturation density, $\rho_0$.  Such a behaviour for the 
kaons was also observed for the zero magnetic field
\cite{isoamss2}. For isospin symmetric nuclear
matter, the increase in the masses of the kaons is due 
to the increase due to the Weinberg Tomozawa term and the first range term,
which dominates over the drop due to the scalar exchange and the 
last two range terms. In the presence of magnetic field,
there is an additional positive shift in the mass of
$K^+$ (given by equation (\ref{meffkpkm})), 
as can be seen in figure \ref{mkp_mag}
as well as from Table I, where the in-medium masses
of $K^+$ have been given for different magnetic fields
((a) without and (b) with the AMM effects)
both for isospin symmetric nuclear matter, as well as, for
asymmetric nuclear matter with $\eta$=0.5. 
The effects of anomalous magnetic moments (AMM) of the
nucleons are observed to have larger masses for
both $K^+$ and $K^0$, as compared to when these 
are not taken into account, as can be observed form the 
figures \ref{mkp_mag} and \ref{mk0_mag}, as well as,
seen from Tables I and II. 
The values of these kaon masses at different magnetic fields
can be compared with the values for zero magnetic field.
As can be seen for the neutral kaon $K^0$ from Table II, 
the mass is observed to increase by maximum value of
about 1 (6) MeV at $\rho_B=\rho_0$ from the zero magnetic field case,
for the symmetric (asymmetric) nuclear matter,
for the magnetic fields considered in the present investigation,
and by a maximum increase of about 13 (25) MeV at $\rho_B=4\rho_0$.
The neutral $K^0$  meson has been studied recently
in the presence of magnetic field for electrically charge
neutral matter, using the chiral SU(3) 
model as used in the present investigation and including one loop 
corrections to the kaon propagator \cite{aguirre_kphi}. 
There is observed to be a similar rise in the mass of $K^0$
as has been obtained in the present work. 

The isospin asymmetry contribution to the mass of the $K^+$ 
($K^0$) is negative (positive) from the Weinberg Tomozawa term.
This leads to a smaller (larger) modification to the
mass of the $K^+$ ($K^0$) mass, due to the vectorial
interaction with the nucleons. The contributions 
from the $\delta$ term in the scalar exchange and the first range
term are positive (negative) for $K^+$ meson 
and of the opposite sign for $K^0$ meson. 
The contribution from $\delta$ term
on the mass of the $K^+$ ($K^0$) in the isospin asymmetric medium 
turns out to be negative (positive) as the contribution from the
first range term dominates over the scalar exchange term.
The drop due to the last term ($d_2$ term) is lessened (enhanced)
due to the isospin asymmetric part, as can be seen from the
expression for the self energy for kaon doublet, given 
by equation (\ref{selfk}). 
The isospin asymmetry is observed to diminish (enhance) the increase
in the mass of the $K^+$ ($K^0$) predominantly due to the
contributions from the Weinberg Tomozawa and the first range term.
These are observed in figures \ref{mkp_mag} and \ref{mk0_mag},
as strong dependence of the $K^0$ mass on isospin asymmetry,
whereas the mass of $K^+$ is observed to be rather insensitive 
to the isospin asymmetry. These can also be seen from the values 
of the masses of the kaon doublet given in Tables I and II, where  
the contribution from the lowest Landau level 
(due to direct interaction with the electromagnetic 
field), in mass of the charged $K^+$, has been taken into account. 
The antikaons are observed to have large drop in isospin 
symmetric nuclear matter, in their
masses, as can be observed from figures \ref{mkm_mag} anf \ref{mk0bar_mag},
as well as from their values as listed in Tables III and IV.
This is due to the attractive interactions of
the Weinberg-Tomozawa term, the scalar exchange 
term as well as from the last two range terms, which dominate 
over the repulsive interaction of the first range term. 
The isospin asymmetry effects lead to larger (smaller) 
masses for the $K^-$ ($\bar {K^0}$) due to opposing effects
arising from the vectorial, scalar exchage and the first
and last range terms of the self energy for the antikaons. 
The isospin effects are observed to show monotonous
behaviour for the masses of the antikaons, as well as for $K^0$. 
However, the isospin effects on the $K^+$ mass
is observed to be rather small and are observed not to follow 
a monotonic behaviour as the other mesons (antikaons and $K^0$
meson), in the present investigation.

The values of the masses of the antikaons 
($K^-$ and $\bar {K^0}$) 
have positive shifts in the presence of the magnetic field
as compared to zero magnetic field. 
In isospin symmetric nuclear matter, 
for $eB=2 m_\pi^2$ $(8 m_\pi^2)$, the value of
$K^-$ mass undergoes a modification from the value of 452 MeV 
at zero magnetic field to 497 (600.7) MeV at $\rho_B=\rho_0$ and
365.8 (501.3) MeV at $\rho_B=4\rho_0$, when
the AMM effects  of nucleons are taken into consideration. 
The isospin asymmetry is observed to lead
to larger values of the $K^-$ mass as compared to the 
isospin symmetric case, and the mass shifts 
for $\eta$=0.5 (as compared to $\eta$=0), 
are upto a maximum value of 
around 15 MeV at $\rho_B=\rho_0$ and 
around 42 MeV at $\rho_B=4\rho_0$ for the values
of the magnetic fields
considered in the present investigation. 
The effects of isospin 
asymmetry are thus observed to be quite negligible,
compared to the medium modifications of $K^-$ mass
with baryon density. The values of the
$K^-$ are smaller when the AMM effects of 
the nucleons are not taken into account and
the modifications arising from the AMM effects 
are observed to be marginal. 
For $\bar{K^0}$, the mass modification is an increase 
upto a maximum value of about 2 (6) MeV from the
zero magnetic field case at $\rho_B=\rho_0$,
and about 6 (12) MeV at $\rho_B=4\rho_0$ for 
$\eta=0 (0.5)$ for the values of magnetic field
considered in the present investigation.
The effect of the baryon density on the masses of the
antikaons (plotted in figures \ref{mkm_mag} and \ref{mk0bar_mag}),
is thus observed to be the most dominant effect as compared 
to the effects from the isospin asymmetry and AMMs of the nucleons. 

The effective masses of the kaons and antikaons, arise
from the values of the scalar fields, $\sigma$, $\zeta$
and $\delta$, which are solved to obtain the values
of the scalar densities of the nucleons. The charged mesons,
in addition, have a positive mass shift, due to interaction
with the electromagnetic field. The modifications of the scalar mesons
are crucial for the medium modifications of the kaons
and anikaons in the magnetized nuclear matter.
The effects of anomalous magnetic moments (AMM) of the
nucleons are incorporated in the expressions for the
scalar and number densities of the nucleons,
while solving the scalar fields. These effects are 
observed to be more dominant for the 
$\sigma$ and $\zeta$ fields, as compared to for the
isovector scalar $\delta$ field as has been studied
in Ref. \cite{dmeson_mag}.
The magnitudes of the scalar fields are observed to be
smaller (leading to larger values for
the fluctuations in these fields from the 
vacuum and hence the contributions from these terms
are larger in magnitude) when the AMM effects are not 
taken into account, as compared to when these
effects are considered. Hence,
the AMM effects are observed to lead to smaller masses
for the kaons and antikaons, as compared to when
these effects are not considered.
The AMM effects are most prominently observed in the
in-medium mass of $K^0$, plotted in figure \ref{mk0_mag}, 
where, the magnitudes in the modifications of the masses
are observed to be of similar order for the
values of isospin asymmetry parameters 
($\eta$=0, 0.3 and 0.5) as shown in the figure.
This is because the modifications of the values
of  $\sigma$ and $\zeta$ are much larger  
as compared to the value of $\delta$,
due to AMM effects \cite{dmeson_mag}.
However, the mass modifications of $K^0$ meson, 
due to AMM effects are observed to 
be smaller when the isospin asymmetry 
of the medium is increased, as can be also seen from their 
values given in Table II as well as from figure \ref{mk0_mag}. 
The AMM effects for the mass of $K^0$ meson,
are more dominant for the higher 
values of the magnetic field ($eB=6m_\pi^2$ and 
$eB=8m_\pi^2$, shown in (c) and (d) 
of figure \ref{mk0_mag}), as expected.

\section{Summary}
To summarize, we have calculated the in-medium masses
of the kaons and antikaons in asymmetric nuclear matter
in the presence of an external magnetic field, using a chiral
SU(3) model. The modifications
of the masses of these strange pseudoscalar mesons arise 
due to the leading and next to leading order terms 
within the chiral perturbative theory. 
The leading contribution (the Weinberg Tomozawa term) 
gives a drop (rise) of the antikaons (kaons) 
due to the attractive 
(repulsive) interaction in the nuclear matter.
The next to leading contributions are due to the
scalar exchange and the range terms.
The effects of the anomalous magnetic moments
(AMM) of the nucleons are taken into consideration
in the present work. 
There is observed to be rise (drop) in the in-medium 
masses of the kaons and the antikaons with density
for given isospin asymmetry and magetic fields.
For the charged $K^\pm$ mesons, there are 
additionl positive shifts in the mass 
in the  presence of magnetic field,
from the lowest Landau level in the present 
investigation. The isospin effetcs  
are observed to be large 
at high densities, and the effect is observed to be
much more prominent for $K^0$ mass as compared to the
antikaons, and $K^+$ masses.
The AMM effcts are observed to be significant for the
mass of $K^0$, expecially for higher values of the
magnetic fields.
The density effects are, however, observed to be 
the dominant medium effects as compared to the effects 
from the magnetic field and isospin asymmetry 
of the magnetized nuclear medium. The mass modification of
$K^0$ meson observed to have large isospin asymmetry
as well as AMM effects, especially
at high densities, as compared to these effects
for the $K^+$, as well as, $K^-$ and $\bar {K^0}$
mesons, should have observable consequences
on the $K^+/K^0$ ratio in the high energy asymmetric  
nuclear collisions.

One of the authors (AM) is grateful to ITP, University of Frankfurt,
for warm hospitality and acknowledges financial support from Alexander
von Humboldt foundation when this work was initiated.

\end{document}